\ifx\mnmacrosloaded\undefined \input mn\fi

\newif\ifAMStwofonts

\ifCUPmtplainloaded \else
  \NewTextAlphabet{textbfit} {cmbxti10} {}
  \NewTextAlphabet{textbfss} {cmssbx10} {}
  \NewMathAlphabet{mathbfit} {cmbxti10} {} 
  \NewMathAlphabet{mathbfss} {cmssbx10} {} 
  \ifAMStwofonts
    \NewSymbolFont{upmath} {eurm10}
    \NewSymbolFont{AMSa} {msam10}
    \NewMathSymbol{\upi}     {0}{upmath}{19}
    \NewMathSymbol{\umu}     {0}{upmath}{16}
    \NewMathSymbol{\upartial}{0}{upmath}{40}
    \NewMathSymbol{\leqslant}{3}{AMSa}{36}
    \NewMathSymbol{\geqslant}{3}{AMSa}{3E}

     \let\le=\leqslant
     
  \else
    \def\umu{\mu}
    \def\upi{\pi}
    \def\upartial{\partial}
  \fi
\fi


\pageoffset{-2.5pc}{0pc}

\loadboldmathnames



\pagerange{1--9}    
\pubyear{1999}
\volume{xxx}

\begintopmatter  

\title{Galactic Globular Clusters as a test for Very Low-Mass stellar models.}

\author{S. Cassisi$^{1,2}$, V. Castellani$^{1,3}$, P. Ciarcelluti$^{1,4}$, 
G. Piotto$^{5}$ and M. Zoccali$^{5}$}

\affiliation{$^1$ Osservatorio Astronomico di Collurania, Via M. Maggini,
 I-64100, Teramo, Italy - E-Mail: cassisi@astrte.te.astro.it}
\smallskip
\affiliation{$^2$ Max-Planck-Institut f\"ur Astrophysik, Karl-Schwarzschild-Strasse 
1, D-85740 Garching, Germany}
\smallskip
\affiliation{$^3$ Universit\'a degli Studi di Pisa, Dipartimento di Fisica,
Piazza Torricelli, I-56100, Pisa, Italy; vittorio@astr18pi.difi.unipi.it}
\smallskip
\affiliation{$^4$ Universit\'a degli studi de L'Aquila, Dipartimento di
Fisica, Via Vetoio, I-67100, L'Aquila, Italy}
\smallskip
\affiliation{$^5$ Universit\`a di Padova, Dipartimento di Astronomia, Vicolo 
dell'Osservatorio 5, 35122 Padova, Italy; piotto@pd.astro.it;manu@obelix.pd.astro.it}

\shortauthor{S. Cassisi et al.}
\shorttitle{Globular cluster VLM stars.}


\acceptedline{}

\abstract{
We make use of the Next Generation model atmospheres by Allard et
al. (1997) and Hauschildt, Allard \& Baron (1999) to compute
theoretical models for low and very low-mass stars for selected
metallicities in the range Z= 0.0002 to 0.002. On this basis, we
present theoretical predictions covering the sequence of H-burning
stars as observed in galactic globulars from the faint end of the Main
Sequence up to, and beyond, the cluster Turn Off. The role played by
the new model atmospheres is discussed, showing that present models
appear in excellent agreement with models by Baraffe et al. (1997) as
computed on quite similar physical basis. One finds that the
theoretical mass - luminosity relations based on this updated set of
models, are in good agreement with the empirical data provided
by Henry \& McCarthy (1993).  Comparison with HST observation
discloses that the location in the Color-Magnitude diagram of
the lower Main Sequence in galactic globular clusters appears again 
in good agreement with the predicted sensitive dependence of these 
sequences on the cluster metallicity.  }


\maketitle  


\section{Introduction}

Very Low Mass (VLM) stars play a relevant role in a wide variety of
astrophysical problems, ranging from the stellar formation processes
to the stellar interior physics and from the Galaxy formation to the
cosmological Dark Matter enigma.  Therefore, in the last decade large
efforts have been devoted to increase the wealth of low-mass and VLM
stars observations both in the visual and near-infrared bands.  A
substantial progress in this field has been recently achieved thanks
to the HST observations which disclosed in several galactic globular
clusters (GCs) a beautiful sequence of central H-burning stars (Piotto
et al. 1997, Ferraro et al. 1997, Marconi et al. 1998), sometimes
nearly down to the lower mass limit for hydrogen burning structures
(King et al. 1998).

The theoretical modeling of VLM stars has been for a long time quite a
difficult task, due to the high densities and the low
temperatures which characterize the structure of these stars. As a
consequence, reliable theoretical predictions on both effective
temperature and luminosity, as needed to understand the Color-Magnitude
(CM) diagram of VLM stars, have been for long time severely challenged.  
However, in the last years the situation has significantly improved thanks to the
relevant progress both in the physics of the stellar interiors as well
as in the treatment of the stellar atmospheric layers (for a detailed
review on this topic see Allard et al. 1997).  As a result,
theoretical investigations based on this new physical framework have
been already able to achieve a satisfactory agreement with several
observational evidences (see Kroupa \& Tout 1997, Baraffe et al. 1997,
hereinafter BCAH97, Chabrier \& Baraffe 1997, Brocato, 
Cassisi \& Castellani 1998, hereinafter BCC98, and references therein).

This is the third of a series of papers (see Alexander et al. 1997 and
BCC98) devoted to investigate VLM structures: in the first paper
(Alexander et al. 1998), we have investigated the effect of using the most
updated physical inputs - as far as it concerns the equation of state
and the low-temperature opacity - on the theoretical prescriptions for
VLM stars; the second one (BCC98) analyzed the effects
on the evolutionary models of the different approaches adopted for
fixing the outer boundary conditions in stellar structure
computations.

In the present work, we will take advantage of the recent availability of
the "Next Generation" (NG) model atmospheres provided by Allard et
al. (1997) and Hauschildt et al. (1999) to discuss the effects of such
new improved boundary conditions on stellar models for VLM
structures. We will show that the structure of theoretical models
appears scarcely affected by the choice about these
conditions. However, this is not the case for the predicted colours,
which strongly depend on the adopted relation connecting the
theoretical quantities, as luminosity (L) and effective temperature
($T_e$) to the observational ones as magnitudes and colours.  Not
surprisingly, present models, as computed on the
basis of the NG model atmospheres, appear to be in excellent agreement with
the result of similar computations given by BCAH97.
\beginfigure*{1}
\vskip 90mm
\caption{{\bf Figure 1.} The H-R diagram location of present VLM models 
with mass $\le 0.6M_\odot$ and for an age of 10Gyr, as compared with
the models  by Baraffe et al. (1997) or Brocato et al. (1998)
for the labeled assumptions on the metallicity. Dotted lines shows
models computed by adopting the same input physics used in this work
but a gray approximation ($T(\tau)$ relation) in the treatment of the outer 
layers.}
\endfigure
We will also take advantage of the growing number of detailed VLM
sequences in galactic GCs provided by HST to extend the
comparison from NGC6397 to other, now well observed, GCs
disclosing that observational data are largely supporting theoretical
predictions concerning the dependence on the metallicity of the CM
diagram location of VLM sequences.  The layout of this paper is as
follows: as a preliminary step, in the next section we present updated
VLM models, discussing to some extent these results to the light of
the current uncertainties on the mixing length parameter and/or the
cluster age.  On this basis, section 3 will be devoted to a comparison
with observations, pointing out the use of the VLM Main Sequence (MS) as
possible calibrator of the cluster metallicity and/or distance
modulus.  Conclusions and final remarks will close the paper.

\section{The stellar models}

As in the previous papers (Alexander et al. 1997, and BCC98) our
models rely on the equation of state by Saumon et al. (1995) for dense
and cool matter, on low-temperature opacities by Alexander \& Ferguson
(1994) and high-temperature opacities by Rogers \& Iglesias (1992). As
already quoted, in BCC98 the outer boundary conditions were fixed by
adopting model atmospheres by Brett (1995a,b). In this paper we will
follow BCAH97 in using NG model atmospheres, which represent a
significant improvement in this field, due to the use of an updated
treatment of pressure broadening and of molecular line absorption
coefficients (see Allard et al. 1997, and Hauschildt et al. 1999 for
more details). As in BCC98, the basis of the atmosphere has been fixed
at Rosseland optical depth $\tau = 100$, i.e., at a depth large enough
for the diffusion approximation to be valid (see, e.g., Mihalas 1978,
Morel et al. 1994), allowing a safe comparison with the results of
different authors.

VLM models have been computed for selected metallicity values, as given 
by Z=0.0002, 0.0006, 0.001 and 0.002, by adopting in all cases an original 
helium abundance $Y=0.23$.  In order to cover with a homogeneous and
self-consistent theoretical scenario the upper portion of the MS and
the cluster Turn Off (TO), numerical computations have been extended
beyond the proper VLM range, up to $M\sim0.8M_\odot$, adopting for
masses larger than $0.6M_\odot$ the OPAL EOS (Rogers, Swenson \&
Iglesias 1996) which allow the required fine match with the VLM
sequence (see Brocato, Cassisi \& Castellani 1997 for a discussion on
this matter).  Tables 1 to 4 provide relevant informations about the
models, as computed assuming a mixing length parameter $ml = 1.8H_P$
and a cluster age of 10 Gyr. According to the solar bolometric
correction adopted by Allard et al. (1997), we used
$M_{Bol,\odot}=4.73$ mag to derive {\sl VRI} magnitudes in the
Johnson-Cousins system and {\sl JHK} ones in the CIT system. In
addition, we provide the stellar magnitudes in some selected HST
filters, including the NICMOS wide filters F110W, F160W and F807W.
\begintable*{1}
\caption{{\bf Table 1.} Mass, luminosity, effective temperature, 
absolute visual magnitude, colours and magnitudes in selected HST 
filters for stellar models with metallicity Z=0.0002 and Y=0.23 
and for an age 10 Gyr.}
\halign{%
\rm#\hfil&
\hskip1pt\hfil\rm#\hfil&\hskip3pt\hfil\rm#\hfil&\hskip3pt\hfil\rm#\hfil&\hskip3pt\hfil\rm#\hfil&
\hskip3pt\hfil\rm#\hfil&\hskip3pt\hfil\rm#\hfil& \hskip3pt\hfil\rm\hfil#\hfil&
\hskip3pt\hfil\rm\hfil#\hfil&\hskip3pt\hfil\rm\hfil#\hfil&\hskip3pt\hfil\rm\hfil#\hfil&
\hskip3pt\hfil\rm\hfil#\hfil&\hskip3pt\hfil\rm\hfil#\hfil&\hskip3pt\hfil\rm\hfil#\hfil&
\hskip3pt\hfil\rm\hfil#\hfil\cr
$M/M_\odot$ & $\log{L/L_{\odot}}$ & $\log{T_e}$ & $M_V$ & $(V-I)$ & $(V-R)$ & $(V-K)$ & $(I-J)$ &
$(J-H)$& $M_{555}$  & $M_{606}$ &$M_{814}$ &$M_{110W}$ &$M_{160W}$&
$M_{807W}$\cr \noalign{\vskip 10pt}
   .095& -3.182 & 3.473& 14.771 & 2.801 & 1.331 & 4.063 & 1.200 & .154& 14.725& 14.095& 11.960& 11.077& 10.536& 10.681\cr
   .096& -3.137 & 3.481& 14.558 & 2.688 & 1.277 & 4.012 & 1.193 & .194& 14.508& 13.900& 11.857& 10.989& 10.412& 10.526\cr
   .097& -3.095 & 3.488& 14.373 & 2.596 & 1.236 & 3.978 & 1.186 & .229& 14.322& 13.729& 11.760& 10.907& 10.299& 10.384\cr
   .098& -3.057 & 3.495& 14.198 & 2.507 & 1.197 & 3.939 & 1.177 & .261& 14.148& 13.569& 11.673& 10.833& 10.199& 10.257\cr
   .099& -3.021 & 3.501& 14.040 & 2.429 & 1.163 & 3.899 & 1.170 & .288& 13.988& 13.423& 11.589& 10.761& 10.104& 10.143\cr
   .100& -2.989 & 3.506& 13.902 & 2.365 & 1.137 & 3.863 & 1.163 & .310& 13.850& 13.296& 11.515& 10.696& 10.021& 10.045\cr
   .101& -2.959 & 3.511& 13.772 & 2.303 & 1.111 & 3.827 & 1.156 & .330& 13.719& 13.173& 11.446& 10.636&  9.945&  9.953\cr
   .102& -2.931 & 3.515& 13.658 & 2.253 & 1.092 & 3.798 & 1.149 & .347& 13.604& 13.065& 11.380& 10.578&  9.874&  9.871\cr
   .104& -2.882 & 3.522& 13.461 & 2.170 & 1.061 & 3.746 & 1.137 & .373& 13.406& 12.876& 11.265& 10.476&  9.753&  9.730\cr
   .107& -2.819 & 3.530& 13.223 & 2.081 & 1.031 & 3.685 & 1.122 & .399& 13.167& 12.644& 11.114& 10.341&  9.599&  9.557\cr
   .110& -2.765 & 3.537& 13.022 & 2.009 & 1.009 & 3.630 & 1.107 & .419& 12.964& 12.446& 10.984& 10.224&  9.470&  9.413\cr
   .112& -2.734 & 3.541& 12.909 & 1.970 &  .998 & 3.599 & 1.098 & .429& 12.850& 12.334& 10.909& 10.156&  9.397&  9.332\cr
   .115& -2.690 & 3.546& 12.757 & 1.924 &  .986 & 3.562 & 1.087 & .440& 12.697& 12.182& 10.801& 10.059&  9.293&  9.219\cr
   .120& -2.623 & 3.553& 12.533 & 1.864 &  .971 & 3.509 & 1.071 & .453& 12.474& 11.957& 10.637&  9.909&  9.137&  9.049\cr
   .125& -2.566 & 3.559& 12.344 & 1.815 &  .958 & 3.462 & 1.055 & .462& 12.285& 11.768& 10.495&  9.781&  9.005&  8.907\cr
   .130& -2.515 & 3.564& 12.177 & 1.775 &  .947 & 3.420 & 1.041 & .469& 12.119& 11.602& 10.368&  9.664&  8.888&  8.783\cr
   .140& -2.426 & 3.572& 11.894 & 1.715 &  .927 & 3.353 & 1.018 & .477& 11.836& 11.324& 10.144&  9.459&  8.683&  8.566\cr
   .150& -2.347 & 3.578& 11.651 & 1.670 &  .908 & 3.303 & 1.001 & .482& 11.595& 11.089&  9.946&  9.275&  8.501&  8.374\cr
   .160& -2.277 & 3.584& 11.432 & 1.627 &  .887 & 3.250 &  .983 & .485& 11.377& 10.880&  9.768&  9.112&  8.342&  8.207\cr
   .180& -2.153 & 3.592& 11.063 & 1.571 &  .857 & 3.179 &  .958 & .487& 11.010& 10.527&  9.454&  8.816&  8.053&  7.910\cr
   .200& -2.045 & 3.598& 10.749 & 1.530 &  .833 & 3.123 &  .939 & .488& 10.697& 10.226&  9.181&  8.556&  7.798&  7.651\cr
   .220& -1.953 & 3.603& 10.481 & 1.495 &  .813 & 3.074 &  .922 & .488& 10.432&  9.971&  8.947&  8.335&  7.582&  7.433\cr
   .250& -1.836 & 3.609& 10.144 & 1.454 &  .788 & 3.011 &  .903 & .487& 10.096&  9.650&  8.651&  8.054&  7.308&  7.156\cr
   .300& -1.682 & 3.617&  9.701 & 1.400 &  .755 & 2.925 &  .876 & .483&  9.656&  9.226&  8.261&  7.683&  6.948&  6.796\cr
   .350& -1.560 & 3.623&  9.355 & 1.361 &  .731 & 2.858 &  .856 & .480&  9.310&  8.893&  7.952&  7.389&  6.662&  6.512\cr
   .400& -1.444 & 3.629&  9.022 & 1.322 &  .708 & 2.790 &  .837 & .476&  8.980&  8.575&  7.658&  7.110&  6.392&  6.245\cr
   .450& -1.298 & 3.640&  8.585 & 1.253 &  .667 & 2.667 &  .802 & .468&  8.544&  8.162&  7.288&  6.765&  6.065&  5.924\cr
   .500& -1.123 & 3.656&  8.051 & 1.159 &  .611 & 2.491 &  .753 & .452&  8.012&  7.661&  6.848&  6.361&  5.686&  5.557\cr
   .600&  -.693 & 3.719&  6.737 &  .868 &  .447 & 1.871 &  .568 & .363&  6.696&  6.438&  5.818&  5.458&  4.918&  4.837\cr
   .650&  -.486 & 3.751&  6.172 &  .750 &  .383 & 1.604 &  .487 & .320&  6.128&  5.906&  5.368&  5.064&  4.587&  4.528\cr
   .700&  -.277 & 3.778&  5.630 &  .657 &  .334 & 1.393 &  .423 & .284&  5.583&  5.389&  4.916&  4.655&  4.231&  4.188\cr
   .800&   .227 & 3.826&  4.340 &  .489 &  .248 & 1.010 &  .311 & .200&  4.288&  4.144&  3.788&  3.600&  3.291&  3.273\cr}
\endtable
\begintable*{2}
\caption{{\bf Table 2.} As Table 1 but for stellar models with metallicity Z=0.0006.}
\halign{%
\rm#\hfil&
\hskip1pt\hfil\rm#\hfil&\hskip3pt\hfil\rm#\hfil&\hskip3pt\hfil\rm#\hfil&\hskip3pt\hfil\rm#\hfil&
\hskip3pt\hfil\rm#\hfil&\hskip3pt\hfil\rm#\hfil& \hskip3pt\hfil\rm\hfil#\hfil&
\hskip3pt\hfil\rm\hfil#\hfil&\hskip3pt\hfil\rm\hfil#\hfil&\hskip3pt\hfil\rm\hfil#\hfil&
\hskip3pt\hfil\rm\hfil#\hfil&\hskip3pt\hfil\rm\hfil#\hfil&\hskip3pt\hfil\rm\hfil#\hfil&
\hskip3pt\hfil\rm\hfil#\hfil\cr
$M/M_\odot$ & $\log{L/L_{\odot}}$ & $\log{T_e}$ & $M_V$ & $(V-I)$ & $(V-R)$ & $(V-K)$ & $(I-J)$ &
$(J-H)$& $M_{555}$  & $M_{606}$ &$M_{814}$ &$M_{110W}$ &$M_{160W}$&
$M_{807W}$\cr \noalign{\vskip 10pt}
   .0929& -3.292&  3.444& 15.985&  3.550&  1.688&  5.234&  1.479&  .267& 15.956& 15.264& 12.471& 11.346& 10.667& 10.806\cr
   .0930& -3.288&  3.445& 15.955&  3.530&  1.676&  5.217&  1.476&  .271& 15.925& 15.236& 12.459& 11.338& 10.656& 10.791\cr
   .0940& -3.242&  3.453& 15.673&  3.367&  1.581&  5.086&  1.449&  .308& 15.644& 14.983& 12.334& 11.245& 10.533& 10.641\cr
   .0950& -3.199&  3.461& 15.404&  3.209&  1.494&  4.957&  1.421&  .341& 15.374& 14.737& 12.216& 11.160& 10.420& 10.501\cr
   .1000& -3.028&  3.490& 14.434&  2.680&  1.236&  4.510&  1.317&  .446& 14.393& 13.830& 11.755& 10.812&  9.995&  9.988\cr
   .1100& -2.820&  3.518& 13.489&  2.264&  1.073&  4.125&  1.221&  .511& 13.441& 12.920& 11.208& 10.361&  9.505&  9.432\cr
   .1200& -2.686&  3.533& 12.965&  2.084&  1.016&  3.937&  1.169&  .531& 12.914& 12.404& 10.859& 10.058&  9.198&  9.097\cr
   .1300& -2.581&  3.543& 12.592&  1.980&   .987&  3.816&  1.135&  .537& 12.540& 12.031& 10.587&  9.814&  8.959&  8.841\cr
   .1500& -2.409&  3.559& 12.004&  1.833&   .952&  3.631&  1.080&  .535& 11.949& 11.439& 10.141&  9.413&  8.575&  8.434\cr
   .1700& -2.275&  3.570& 11.571&  1.744&   .928&  3.505&  1.042&  .530& 11.516& 11.007&  9.794&  9.096&  8.273&  8.121\cr
   .2000& -2.107&  3.581& 11.057&  1.660&   .898&  3.380&  1.003&  .523& 11.004& 10.504&  9.363&  8.694&  7.889&  7.726\cr
   .2500& -1.897&  3.592& 10.441&  1.579&   .861&  3.250&   .964&  .516& 10.389&  9.905&  8.825&  8.186&  7.398&  7.231\cr
   .3000& -1.741&  3.599&  9.996&  1.531&   .835&  3.169&   .940&  .514&  9.944&  9.473&  8.428&  7.806&  7.027&  6.860\cr
   .3500& -1.618&  3.605&  9.642&  1.489&   .812&  3.099&   .919&  .512&  9.591&  9.133&  8.114&  7.507&  6.735&  6.569\cr
   .4000& -1.495&  3.612&  9.282&  1.441&   .785&  3.017&   .896&  .510&  9.232&  8.789&  7.801&  7.212&  6.447&  6.283\cr
   .4500& -1.341&  3.623&  8.819&  1.368&   .742&  2.895&   .864&  .511&  8.771&  8.351&  7.411&  6.846&  6.088&  5.931\cr
   .5000& -1.158&  3.640&  8.248&  1.255&   .678&  2.711&   .817&  .512&  8.200&  7.817&  6.949&  6.421&  5.671&  5.528\cr
   .5500&  -.956&  3.668&  7.573&  1.084&   .580&  2.401&   .738&  .482&  7.524&  7.198&  6.442&  5.973&  5.269&  5.148\cr
   .5800&  -.836&  3.688&  7.175&   .982&   .518&  2.175&   .671&  .440&  7.129&  6.834&  6.143&  5.719&  5.073&  4.970\cr
   .6000&  -.724&  3.708&  6.823&   .896&   .467&  1.965&   .604&  .397&  6.779&  6.512&  5.876&  5.496&  4.912&  4.826\cr
   .6500&  -.518&  3.742&  6.236&   .773&   .399&  1.661&   .505&  .335&  6.192&  5.962&  5.410&  5.095&  4.597&  4.536\cr
   .7000&  -.314&  3.769&  5.703&   .682&   .349&  1.447&   .439&  .295&  5.655&  5.454&  4.964&  4.694&  4.254&  4.209\cr
   .8000&   .157&  3.813&  4.509&   .533&   .271&  1.104&   .337&  .228&  4.455&  4.300&  3.915&  3.710&  3.367&  3.346\cr}
\endtable
\begintable*{3}
\caption{{\bf Table 3.} As Table 1 but for stellar models with metallicity Z=0.001.}
\halign{%
\rm#\hfil&
\hskip1pt\hfil\rm#\hfil&\hskip3pt\hfil\rm#\hfil&\hskip3pt\hfil\rm#\hfil&\hskip3pt\hfil\rm#\hfil&
\hskip3pt\hfil\rm#\hfil&\hskip3pt\hfil\rm#\hfil& \hskip3pt\hfil\rm\hfil#\hfil&
\hskip3pt\hfil\rm\hfil#\hfil&\hskip3pt\hfil\rm\hfil#\hfil&\hskip3pt\hfil\rm\hfil#\hfil&
\hskip3pt\hfil\rm\hfil#\hfil&\hskip3pt\hfil\rm\hfil#\hfil&\hskip3pt\hfil\rm\hfil#\hfil&
\hskip3pt\hfil\rm\hfil#\hfil\cr
$M/M_\odot$ & $\log{L/L_{\odot}}$ & $\log{T_e}$ & $M_V$ & $(V-I)$ & $(V-R)$ & $(V-K)$ & $(I-J)$ &
$(J-H)$& $M_{555}$  & $M_{606}$ &$M_{814}$ &$M_{110W}$ &$M_{160W}$&
$M_{807W}$\cr \noalign{\vskip 10pt}
   .0917& -3.354&  3.429& 16.687&  3.956&  1.915&  5.878&  1.663& .306& 16.659& 15.935&  12.781& 11.501& 10.775& 10.921\cr
   .0920& -3.339&  3.432& 16.577&  3.891&  1.872&  5.817&  1.647& .318& 16.550& 15.838&  12.735& 11.470& 10.734& 10.871\cr
   .0930& -3.292&  3.440& 16.267&  3.713&  1.759&  5.654&  1.609& .352& 16.241& 15.561&  12.598& 11.373& 10.609& 10.721\cr
   .0940& -3.249&  3.448& 15.971&  3.541&  1.655&  5.493&  1.569& .383& 15.946& 15.293&  12.471& 11.284& 10.498& 10.585\cr
   .0950& -3.209&  3.455& 15.711&  3.393&  1.570&  5.355&  1.534& .409& 15.685& 15.054&  12.353& 11.201& 10.396& 10.461\cr
   .0960& -3.171&  3.462& 15.458&  3.248&  1.490&  5.217&  1.499& .432& 15.431& 14.820&  12.241& 11.124& 10.301& 10.345\cr
   .0980& -3.103&  3.473& 15.045&  3.022&  1.373&  5.003&  1.446& .467& 15.012& 14.432&  12.045& 10.982& 10.135& 10.147\cr
   .1000& -3.045&  3.483& 14.695&  2.833&  1.283&  4.817&  1.398& .493& 14.659& 14.100&  11.876& 10.861&  9.997&  9.983\cr
   .1050& -2.930&  3.499& 14.112&  2.559&  1.169&  4.542&  1.327& .527& 14.073& 13.540&  11.555& 10.611&  9.728&  9.675\cr
   .1100& -2.843&  3.510& 13.713&  2.391&  1.105&  4.363&  1.281& .544& 13.670& 13.151&  11.317& 10.416&  9.528&  9.452\cr
   .1200& -2.711&  3.524& 13.178&  2.203&  1.041&  4.151&  1.223& .557& 13.132& 12.625&  10.961& 10.114&  9.225&  9.125\cr
   .1300& -2.607&  3.535& 12.777&  2.076&  1.003&  3.996&  1.180& .560& 12.728& 12.225&  10.683&  9.873&  8.992&  8.874\cr
   .1500& -2.438&  3.550& 12.184&  1.923&   .965&  3.794&  1.122& .555& 12.132& 11.628&  10.236&  9.474&  8.611&  8.472\cr
   .2000& -2.134&  3.573& 11.203&  1.727&   .917&  3.503&  1.036& .537& 11.149& 10.644&   9.444&  8.751&  7.926&  7.762\cr
   .2500& -1.921&  3.585& 10.568&  1.637&   .885&  3.358&   .992& .527& 10.515& 10.021&   8.897&  8.236&  7.431&  7.260\cr
   .3000& -1.764&  3.593& 10.111&  1.579&   .859&  3.261&   .964& .521& 10.058&  9.576&   8.495&  7.855&  7.063&  6.890\cr
   .3500& -1.640&  3.598&  9.759&  1.542&   .841&  3.197&   .946& .519&  9.707&  9.234&   8.180&  7.553&  6.768&  6.594\cr
   .4000& -1.516&  3.605&  9.394&  1.493&   .815&  3.112&   .921& .516&  9.343&  8.883&   7.862&  7.254&  6.478&  6.306\cr
   .4500& -1.361&  3.616&  8.929&  1.422&   .777&  2.993&   .888& .518&  8.878&  8.440&   7.467&  6.884&  6.113&  5.948\cr
   .5000& -1.181&  3.632&  8.373&  1.317&   .718&  2.827&   .847& .529&  8.321&  7.918&   7.013&  6.464&  5.691&  5.538\cr
   .5500&  -.979&  3.659&  7.693&  1.140&   .618&  2.530&   .776& .517&  7.639&  7.295&   6.508&  6.014&  5.264&  5.137\cr
   .6000&  -.752&  3.698&  6.923&   .936&   .494&  2.076&   .642& .429&  6.876&  6.596&   5.936&  5.534&  4.906&  4.811\cr
   .6500&  -.549&  3.733&  6.318&   .801&   .417&  1.736&   .530& .353&  6.272&  6.033&   5.463&  5.132&  4.610&  4.540\cr
   .7000&  -.349&  3.761&  5.786&   .704&   .364&  1.502&   .456& .310&  5.739&  5.530&   5.026&  4.744&  4.286&  4.239\cr
   .8000&   .099&  3.804&  4.640&   .560&   .288&  1.167&   .357& .237&  4.588&  4.423&   4.020&  3.804&  3.444&  3.417\cr}
\endtable
\begintable*{4}
\caption{{\bf Table 4.} As Table 1 but for stellar models with metallicity Z=0.002.}
\halign{%
\rm#\hfil&
\hskip1pt\hfil\rm#\hfil&\hskip3pt\hfil\rm#\hfil&\hskip3pt\hfil\rm#\hfil&\hskip3pt\hfil\rm#\hfil&
\hskip3pt\hfil\rm#\hfil&\hskip3pt\hfil\rm#\hfil& \hskip3pt\hfil\rm\hfil#\hfil&
\hskip3pt\hfil\rm\hfil#\hfil&\hskip3pt\hfil\rm\hfil#\hfil&\hskip3pt\hfil\rm\hfil#\hfil&
\hskip3pt\hfil\rm\hfil#\hfil&\hskip3pt\hfil\rm\hfil#\hfil&\hskip3pt\hfil\rm\hfil#\hfil&
\hskip3pt\hfil\rm\hfil#\hfil\cr
$M/M_\odot$ & $\log{L/L_{\odot}}$ & $\log{T_e}$ & $M_V$ & $(V-I)$ & $(V-R)$ & $(V-K)$ & $(I-J)$ &
$(J-H)$& $M_{555}$  & $M_{606}$ &$M_{814}$ &$M_{110W}$ &$M_{160W}$&
$M_{807W}$\cr \noalign{\vskip 10pt}
   .0894& -3.473&  3.401& 17.982&  4.596&  2.287&  7.028&  2.094& .363& 17.952& 17.149& 13.436& 11.809& 11.008& 11.198\cr
   .0895& -3.468&  3.402& 17.943&  4.576&  2.274&  7.004&  2.086& .366& 17.914& 17.115& 13.418& 11.798& 10.995& 11.182\cr
   .0897& -3.457&  3.404& 17.864&  4.535&  2.246&  6.957&  2.069& .373& 17.835& 17.047& 13.379& 11.773& 10.965& 11.148\cr
   .0900& -3.441&  3.407& 17.746&  4.473&  2.204&  6.886&  2.045& .384& 17.718& 16.945& 13.323& 11.738& 10.923& 11.098\cr
   .0910& -3.392&  3.416& 17.388&  4.285&  2.077&  6.672&  1.974& .414& 17.363& 16.633& 13.152& 11.630& 10.794& 10.946\cr
   .0920& -3.346&  3.425& 17.037&  4.094&  1.951&  6.458&  1.904& .442& 17.014& 16.323& 12.992& 11.530& 10.675& 10.802\cr
   .0930& -3.304&  3.432& 16.746&  3.940&  1.851&  6.290&  1.851& .463& 16.724& 16.062& 12.853& 11.437& 10.568& 10.675\cr
   .0950& -3.227&  3.446& 16.186&  3.634&  1.663&  5.954&  1.747& .500& 16.165& 15.551& 12.595& 11.271& 10.377& 10.443\cr
   .0970& -3.159&  3.458& 15.712&  3.380&  1.516&  5.670&  1.661& .526& 15.690& 15.112& 12.371& 11.123& 10.214& 10.246\cr
   .1000& -3.074&  3.471& 15.176&  3.106&  1.371&  5.362&  1.572& .551& 15.151& 14.607& 12.103& 10.935& 10.014& 10.012\cr
   .1100& -2.882&  3.497& 14.127&  2.623&  1.161&  4.789&  1.406& .581& 14.094& 13.586& 11.520& 10.503&  9.575&  9.512\cr
   .1200& -2.752&  3.511& 13.545&  2.403&  1.081&  4.515&  1.328& .587& 13.507& 13.010& 11.147& 10.201&  9.280&  9.189\cr
   .1300& -2.649&  3.521& 13.125&  2.265&  1.037&  4.333&  1.276& .586& 13.084& 12.593& 10.859&  9.960&  9.047&  8.939\cr
   .1500& -2.483&  3.536& 12.499&  2.087&   .987&  4.085&  1.204& .580& 12.454& 11.966& 10.402&  9.566&  8.672&  8.542\cr
   .1700& -2.349&  3.547& 12.029&  1.973&   .960&  3.915&  1.154& .570& 11.981& 11.491& 10.040&  9.247&  8.371&  8.227\cr
   .2000& -2.177&  3.560& 11.454&  1.849&   .935&  3.721&  1.097& .555& 11.403& 10.907&  9.582&  8.836&  7.986&  7.825\cr
   .2500& -1.964&  3.573& 10.794&  1.739&   .911&  3.541&  1.044& .542& 10.742& 10.245&  9.026&  8.324&  7.495&  7.322\cr
   .3000& -1.804&  3.581& 10.321&  1.675&   .894&  3.435&  1.012& .535& 10.269&  9.775&  8.613&  7.937&  7.121&  6.943\cr
   .3500& -1.678&  3.587&  9.955&  1.631&   .878&  3.357&   .989& .532&  9.902&  9.413&  8.290&  7.632&  6.824&  6.644\cr
   .4000& -1.552&  3.593&  9.591&  1.588&   .860&  3.283&   .967& .530&  9.538&  9.057&  7.968&  7.326&  6.525&  6.344\cr
   .4500& -1.402&  3.602&  9.148&  1.527&   .833&  3.180&   .938& .531&  9.096&  8.628&  7.584&  6.965&  6.167&  5.990\cr
   .5000& -1.223&  3.619&  8.583&  1.415&   .777&  3.004&   .891& .549&  8.529&  8.093&  7.128&  6.547&  5.745&  5.578\cr
   .5500& -1.031&  3.642&  7.954&  1.262&   .698&  2.771&   .839& .564&  7.893&  7.508&  6.647&  6.110&  5.301&  5.154\cr
   .6000&  -.834&  3.674&  7.250&  1.057&   .577&  2.377&   .737& .509&  7.193&  6.874&  6.144&  5.682&  4.947&  4.831\cr
   .6500&  -.611&  3.715&  6.508&   .867&   .458&  1.910&   .590& .399&  6.460&  6.199&  5.588&  5.221&  4.635&  4.553\cr
   .7000&  -.417&  3.744&  5.955&   .760&   .398&  1.638&   .501& .338&  5.907&  5.680&  5.141&  4.831&  4.330&  4.269\cr
   .8000&  -.005&  3.787&  4.884&   .613&   .318&  1.287&   .392& .262&  4.833&  4.652&  4.213&  3.974&  3.579&  3.545\cr}
\endtable

Figure 1 shows the location in the theoretical HR diagram of the new
models with Z=0.0002 and 0.002, as compared with similar models
provided by BCC98, BCAH97 as well as with models computed by 
adopting a $T(\tau)$ relation in the atmospheric layers. As whole,
models in this figure appear all in reasonable agreement; this
occurrence shows that the thermal structures of the models is not
dramatically dependent on the adopted boundaries conditions. As
expected, one finds that the new models based on NG model atmospheres
are virtually identical with those by BCAH97, showing that 
different groups with different evolutionary codes obtain
results which only depend on the adopted input physics. One may
finally notice that Brett's (1995a,b) model atmospheres 
provided a description of the thermal structure of VLM stars which is
comparable with the most updated NG models over a rather large range
of effective temperatures. However, in the region around
$\rm\log{T_e}\sim3.63$ the new models based on the NG model atmospheres 
appear slightly hotter and marginally more luminous than the ones
computed by adopting the boundary conditions given by Brett (1995a,b). 
As discussed by Bessel (1995), such a difference is probably due to 
the too strong $H_2O$ bands in Brett's model atmospheres, which produces 
an overestimate of the $H_2O$ opacity for temperatures cooler then 3000K.
\beginfigure{2}
\vskip 82mm
\caption{{\bf Figure 2.} The predicted
location in the ($M_V, V-I$) CM diagram of VLM sequence from the
present paper (points) and from BCAH97 (solid line) according to
evaluations of colours based on the NG model atmospheres, and for two 
assumptions about the stellar metallicity: Z=0.0002 and 0.002.
Dotted lines show the location of the same theoretical sequences but 
adopting Allard and Hauschildt's (1995) colours (see text).}
\endfigure

However, it is worth noticing that the colour - effective temperature
relations based on different model atmospheres databases, show remarkable
differences. 
This is shown in figure 2 where we compare VLM
sequences in the observational ($M_V, V-I$) diagram for selected
metallicities, as evaluated by adopting alternatively the bolometric
corrections and colour - effective temperature relations provided
either by the NG model atmospheres or by Allard \& Hauschildt's (1995)
(implemented at effective temperatures larger than 4000K with the
Kurucz's (1993) transformations), as in BCC98.  The significant
differences between the two approaches show that whereas model atmospheres 
are already good enough to produce reliable stellar models, the prediction
about colours is still uncertain, affected by even minor details in the
evaluation of predicted spectra. In this field, the NG model atmospheres
represent a significant improvement in comparison with the previous 
theoretical results.

Present theoretical predictions concerning the CMD location of VLM
sequences are reported in figure 3 for the four selected metallicities
and for an assumed cluster age of 10 Gyr. As already known, one finds
that at lower luminosities the VLM sequence is predicted to be a
rather sensitive indicator of the cluster metallicity. As a relevant
point, numerical experiments fully support the BCAH97's finding that
the VLM sequence for magnitudes below $M_V\sim7$ mag is expected to be
strictly independent of assumptions about either the cluster age or
the efficiency of super-adiabatic convection. One concludes that the
run of VLM sequences in the CM diagram is among the "not-too-abundant"
observational quantities firmly predicted by theory, only depending on
the reliability of the adopted physical scenario.
\beginfigure{3}
\vskip 82mm
\caption{{\bf Figure 3.} Comparison between theoretical isochrones 
in the ($M_V, V-I$) diagram for an age 10Gyr and for the labeled assumptions 
about the metallicity.}
\endfigure

Figure 4 finally gives the mass - luminosity relations for all the
investigated metallicities and for selected photometric bands ranging
from the visual to the near-infrared. It is worth noticing (see also
BCC98) that the mass - luminosity relation becomes more and more
insensitive to the stellar metallicity when going from the visual to
the near-infrared photometric bands, with the $m - M_K$ relation
almost unaffected by the metallicity.  This occurrence together with
the evidence that in the near-IR the effects of extinction and
differential reddening are considerably reduced, can be of help when
investigating the mass distribution of VLM objects in the field and in
the galactic bulge (see for instance, Zoccali et al. 1999). In the
same figure, the empirical data by Henry \& McCarthy (1993)
for a sample of visual and eclipsing binaries in the solar
neighborhood are also shown.  Since such a sample is expected to cover a significant
spread in metallicity, fig. 4 has been implemented with the
theoretical mass - luminosity relation for solar composition, as
computed by using the same physical inputs adopted for the more metal
poor VLM sequences. One can easily notice that, within the current
uncertainties and the significant dispersion (due to a spread in the
metallicity and, probably also in the age) in the observational data,
there is a quite good agreement between empirical and theoretical mass -
luminosity relations.
\beginfigure{4}
\vskip 120mm
\caption{{\bf Figure 4.} The mass - luminosity relations for
various assumptions on the heavy elements abundance, in selected photometric
bands. The observational data provided by Henry \& McCarthy (1993) are also
shown. The mass - luminosity relation corresponding to solar metallicity
VLM models has been also plotted (see text for more details).}
\endfigure
\beginfigure{5}
\vskip 80mm
\caption{{\bf Figure 5.} The CMDs of NGC6752 and M30 with the CMD of
NGC6752 shifted to overlap the Turn Off of M30 (see text).}
\endfigure
\beginfigure{6}
\vskip 80mm
\caption{{\bf Figure 6.} Comparison between the HST CMD of NGC6397 and the 
theoretical isochrones. The solid lines correspond to the evolutionary prescriptions 
for the metallicity adopted for the cluster, and the dashed lines show the main 
sequence loci for very low-mass structures, for all the other metallicities accounted 
for in the present work. The age of the various isochrones, the adopted distance
modulus and reddening are also labeled.}
\endfigure

\section{The observational test}

BCAH97 were already able to test their prediction to the high quality
CM diagram of NGC6397 (Cool, Piotto \& King 1996). Unfortunately, a
similar precise comparisons with clusters of different metallicities
(M15 and $\omega~Cen$) was hampered by the poorer quality of
photometric data available at that time. However, the sample of VLM
sequences observed by HST in galactic globulars is increased, and now
one finds five more clusters with sufficiently well defined VLM
sequences, as given by M92, M30 (Piotto et
al. 1997), NGC6752 (Ferraro et al. 1997), M10 and M55 (Piotto \&
Zoccali 1999). Table 5 gives estimates of [Fe/H] for this sample of
"well observed" clusters as provided by Cohen et al. (1999). The same
table gives the values of the global metallicity [M/H] as obtained by
adopting  the prescription for the relation between [Fe/H], [$\alpha$/Fe], and [M/H]
provided by Salaris, Chieffi \& Straniero (1993) and assuming alternatively
[$\alpha$/Fe]=0.25 ($\rm [M/H]_{0.25}$) and [$\alpha$/Fe]=0.35 ($\rm [M/H]_{0.35}$),
in order to take into account the current uncertainty in the $\alpha-$elements
enhancement in GC stars. Current estimates for distance modulus and reddening taken 
from the compilation of Harris (1996) are also listed.
\beginfigure{7}
\vskip 80mm
\caption{{\bf Figure 7.} As figure 6 but for the cluster M30.}
\endfigure
\begintable*{5}
\caption{{\bf Table 5.} Heavy element abundance and selected 
observational parameters for  globular clusters with "well observed"
VLM sequence.}  
\halign{%
\rm#\hfil& \hskip7pt\hfil\rm#\hfil&\hskip7pt\hfil\rm#\hfil& \hskip7pt\hfil\rm#\hfil&\hskip7pt\hfil\rm#
\hfil& \hskip7pt\hfil\rm#\hfil & \hskip7pt\hfil\rm#\hfil & \hskip7pt\hfil\rm#\hfil & \hskip7pt\hfil\rm\hfil#
\hfil & \hskip7pt\hfil\rm#\hfil & \hskip7pt\hfil\rm#\hfil\cr
NGC & Name &  HST filters  &[Fe/H]$_C$  &  [M/H]$_{0.25}$ & [M/H]$_{0.35}$  & $\rm E(V-I)$ & $(m-M)_V$ \cr
\noalign{\vskip 10pt}
6341  & M92   &F606W, F814W &   -2.10  & -1.93 & -1.85 & 0.03 & 14.64 \cr
7099  & M30   &F555W, F814W &   -1.94  & -1.77 & -1.69 & 0.04 & 14.62 \cr
6397  &       &F555W, F814W &   -1.78  & -1.61 & -1.53 & 0.24 & 12.36 \cr
6809  & M55   &F606W, F814W &   -1.65  & -1.47 & -1.40 & 0.09 & 13.87 \cr
6752  &       &F555W, F814W &   -1.41  & -1.24 & -1.16 & 0.05 & 13.13 \cr
6254  & M10   &F606W, F814W &   -1.38  & -1.21 & -1.13 & 0.38 & 14.08 \cr}
\endtable

However, as shown in the same table, three clusters out of this sample
(M92, M55 and M10) have been observed with the HST F606W filter, so
that data are not homogeneous with NGC6397 and, in addition, their
calibration can be affected by not negligible uncertainties. Thus we
will limit the following discussion only to the three clusters M30,
NGC6397 and NGC6752. The observational data for these clusters have
been translated from the HST photometric system to the standard
Johnson-Cousins system by adopting the Holtzman et al. (1995)
recipes.

As a first crude test of the metal dependence predicted by theory, let
us show in figure 5 the data for NGC6752, the more metal-rich cluster
in our sample, shifted onto the M30 data in such a way that the location of
the Turn Off points in both clusters coincides. Comparison with theoretical
predictions given in figure 3, shows that there is a clear evidence for 
the existence of the effect of metallicity on the location of
the lower main sequences - with a shift toward redder colours when
increasing the metallicity -, which appears in good, even if qualitative, 
agreement with theoretical predictions.

A precise comparison between observations and theoretical predictions
would require an exact knowledge of both the cluster reddening and
distance modulus. As a less stringent but still significative
approach, one can adopt, e.g., a suitable value for the cluster
reddening, regarding the distance modulus as a free parameter to be
determined by best fitting theoretical predictions: the accuracy of
the fitting all along the MS will be a non trivial indicator of the
reliability of the theoretical scenario.  As a first application of
this last procedure, let us follow BCAH97 assuming for NGC6397, $\rm
E(B-V)= 0.18$ which corresponds to $\rm E(V-I)= 0.24$ by adopting the
relation $\rm E(V-I)=1.35\cdot{E(B-V)}$ (Drukier et al. 1993, He et
al.  1995). Not surprisingly, figure 6 shows that a good
fit is achieved adopting [M/H]=$-1.5$ and $(m-M)_0= 11.9$
(corresponding to $(m-M)_V= 12.46$), as already found by BCAH97. The
run of VLM sequences for different metallicities, as given in the same
figure, discloses the good sensitivity of this fitting to the assumed
metal content. One may notice that the derived distance modulus is in
satisfactory agreement with the independent evaluation given by Reid
\& Gizis (1998) on the basis of Hipparcos M-subdwarf Main-Sequence
fitting.

In addition, now one can note that the comparison of the cluster Turn-Off 
luminosity with theoretical isochrones as computed for selected
ages and without allowing for the efficiency of element sedimentation
(see Cassisi et al. 1998, 1999) would assign to the cluster an age of,
about, 12 Gyr, in good agreement with the age recently discussed on
quite an independent way by Salaris \& Weiss (1997). If element
diffusion is at work in Pop.~II globular cluster stars, thus the age
of the cluster would be further reduced by an additional $\sim0.7$ Gyr
(see Castellani et al. 1997, Cassisi et al. 1998, 1999). However, one
has also to bear in mind that the derived age is dependent on the
assumption about the cluster reddening: if the reddening is reduced by 
$\rm \Delta{E(V-I)}\sim0.03$ the age estimated for the cluster is
increased by $\sim1$ Gyr.

One can take advantage of the range of metallicities covered by our
cluster sample to explore theoretical predictions in a rather large
range of metallicities. Figure 7 shows that adopting for M30 $\rm
E(B-V)= 0.03$ and thus $\rm E(V-I)= 0.04$ from the Harris's
compilation, the fitting of the cluster loci by the theoretical
sequence for [M/H]=$-2.0$ requires a distance modulus equal to
$(m-M)_V=14.98$, which is about 0.3 mag larger than the value listed
in the Harris's compilation, but in good agreement with the value
recently given by Gratton et al (1997: $(m-M)_V=14.94\pm0.08$)
obtained from the main-sequence fitting to Hipparcos subdwarfs.  By
comparing the theoretical prescriptions for the TO luminosity with the
cluster CM diagram, one predicts for this cluster again an age of the
order of 12 Gyr.

Figure 8 (top panel) shows the result of the same procedure but for
the cluster NGC6752. Assuming again from Harris data $\rm E(B-V)=
0.04$, i.e. $\rm E(V-I)=0.05$, and adopting a metallicity
[M/H]=$-1.3$, the best fit is achieved for a distance modulus equal to
$(m-M)_V=13.08$, in reasonable agreement with the value given by
Harris (1996) and with the estimate given by Renzini et al. (1996)
from the white dwarfs cooling sequence, but smaller by about 0.1 mag
than the determination by Gratton et al. (1997). From this figure, it
is easy to notice that the observed lower main-sequence location seems
to be in better agreement with the theoretical prescriptions for a
metallicity [M/H]$\sim-1.0$ rather than with [M/H]=$-1.3$. This
occurrence can be due to several reasons: problems in the calibration
of the observational data from the HST photometry to the standard
system and/or drawbacks in the adopted colour - effective
temperature relation for this moderately metal-rich mixture, and/or the
uncertainty in the estimation of the global metallicity suitable for
the cluster (see table 5).
It is also surprising to notice that the fitting given in Figure 8
(top panel) implies a cluster age larger or of the order of 14 Gyr,
and one is reluctant to conclude for such large age. 
In order to reduce this age down to $\sim 12$ Gyr, the cluster reddening
would have to be doubled and this choice is not supported by current
determinations of such a parameter. 
However, the cluster age could be significantly reduced if, as suggested by 
the VLM sequence, we accept for the cluster a metallicity [M/H]$\sim -1.0$ - a value
larger than the one usually adopted for this GC -, since increasing the metallicity
decreases the magnitude of the Turn-Off for each given cluster age.
This has been done in the bottom panel of figure 8, where we compare
the CM diagram of NGC6752 with the theoretical isochrones
corresponding to a stellar metallicity [M/H]$= -1.0$ (i.e. Z=0.002);
now the derived age for the cluster is of the order of 12 Gyr.
When accounting for the quoted uncertainties on both the photometric
data calibration and the theoretical colour - effective temperatures
relation, this result can not be considered as a plain evidence
that NGC6752 has a so large metallicity (see, for instance, Vandenberg 2000). 
However, it is clear that a more reliable investigation on the GCs properties 
requires also a more accurate evaluation of the GC metallicity scale 
(Rutledge, Hesser \& Stetson 1997), and also an homogeneous measurement 
of both the iron content and $\alpha-$elements abundance.

\beginfigure{8}
\vskip 120mm
\caption{{\bf Figure 8.} {\sl Top Panel}: as figure 6 but for the 
cluster NGC6752.
{\sl Bottom Panel}: as top panel but by adopting for the theoretical
framework a metallicity equal to [M/H]$=-1.0$ (i.e. Z=0.002) (see text
for more details).}
\endfigure

\section{Conclusions}

In this paper we have presented new models for metal poor VLM MS stars
based on the most updated physical scenario.  In particular, present
computations rely on the set of model atmospheres for M dwarfs
recently provided by Allard et al. (1997) and Hauschildt et
al. (1999).  The role played by model atmospheres in predicting the
observational properties of VLM stellar models has been discussed by
comparing models computed under different boundary conditions. As a
result, it appears that predictions about the star luminosities and
radii are fairly solid whereas the predicted colours of the stellar models 
are critically dependent on the adopted model atmospheres which
could be still affected by not negligible uncertainties (see below).

As already known, we found that the slope of the VLM sequence is
sensitively dependent on the stellar metallicity.  In addition, we
show that the CMD location of VLM stars observed by HST in selected
galactic globulars is well reproduced by theoretical stellar models.
This occurrence appears as a plain evidence that present theoretical
framework for VLM structures is able to finely reproduce the
dependence of the observational properties of M dwarfs on the heavy
elements abundance.

In this context it is worth noticing that, if and when the cluster
reddening is known, the magnitude of the MS at $\rm (V-I)_0=0.95$ mag,
could be used as a standard candle to derive information on the
cluster distance modulus and, in turn, on the cluster age. In this
respect, we note the small uncertainties on the cluster reddening do
not play a dramatic role. In fact, present evolutionary models predict
for the main sequence slope around $M_V\approx 7$ mag,
$\partial{M_V}/\partial{(V-I)}\sim4.3$.  Therefore, an uncertainty
in the reddening of the order of $\Delta{E(V-I)}\sim0.02$ mag would imply
a corresponding uncertainty in the cluster distance modulus of about
$\Delta(m-M)_V\sim 0.09$ mag. According to current calibrations of the
Turn Off magnitude as a function of the age (see, e.g., Cassisi et
al. 1998, 1999), the quoted uncertainty would to constraint the
cluster age within $\sim\pm1.2$ Gyr.

At larger metallicities, theoretical models predict a CMD location of
the MS which, for each given colour, appear slightly fainter when
compared with the observations. As it has been discussed in the
previous section, this occurrence can be due to a significant
uncertainty in the adopted metallicity, but it can be also a first
signature of a real shortcoming of the models, as shown by Baraffe et
al. (1998), and confirmed by our own computations, by comparing theory
and observations at solar metallicity. Following the early suggestion
provided by Alvarez \& Plez (1998), Baraffe et al. (1998) have shown
by performing some numerical experiments, that such discrepancy
between models and observations could be solved by accounting for a
missing source of opacity for wavelengths shorter than $1{\mu}m$.  It
is worth noticing that an increase in the stellar opacity in this
spectral range would improve the match between theory and observations
in the optical, but without affecting the near-infrared colours. This
occurrence allow us to be confident in the results so far obtained in
comparing VLM models with near-infrared photometric data at solar
metallicity (see for instance, Zoccali et al. 1999).

\section*{Acknowledgments} 

We gratefully acknowledge F. Allard and P.H. Hauschildt for kindly
providing us with their updated "Next Generation" model atmospheres as
well as for useful informations and suggestions about these models.
We are also grateful to an anonymous referee for the pertinence of
her/his comments regarding the content of an early draft of this
paper, which significantly improved its readability.

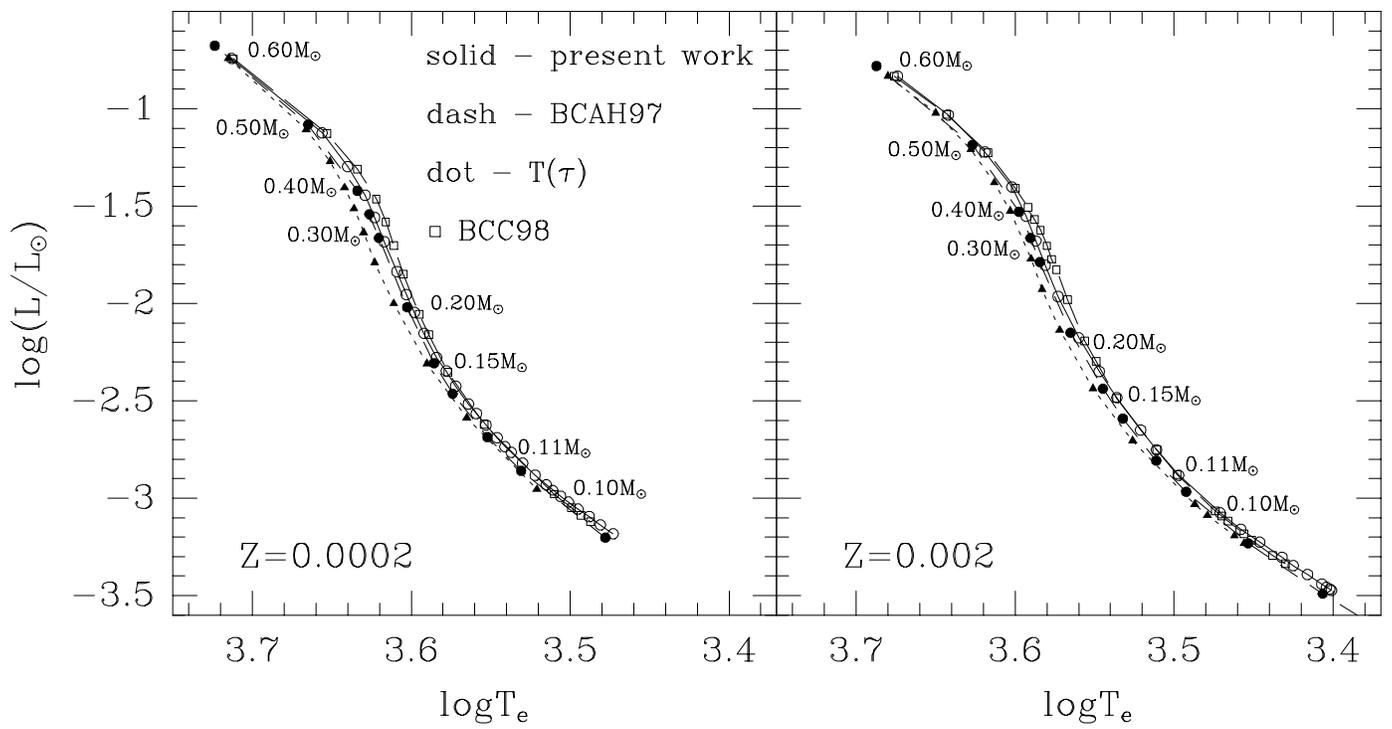
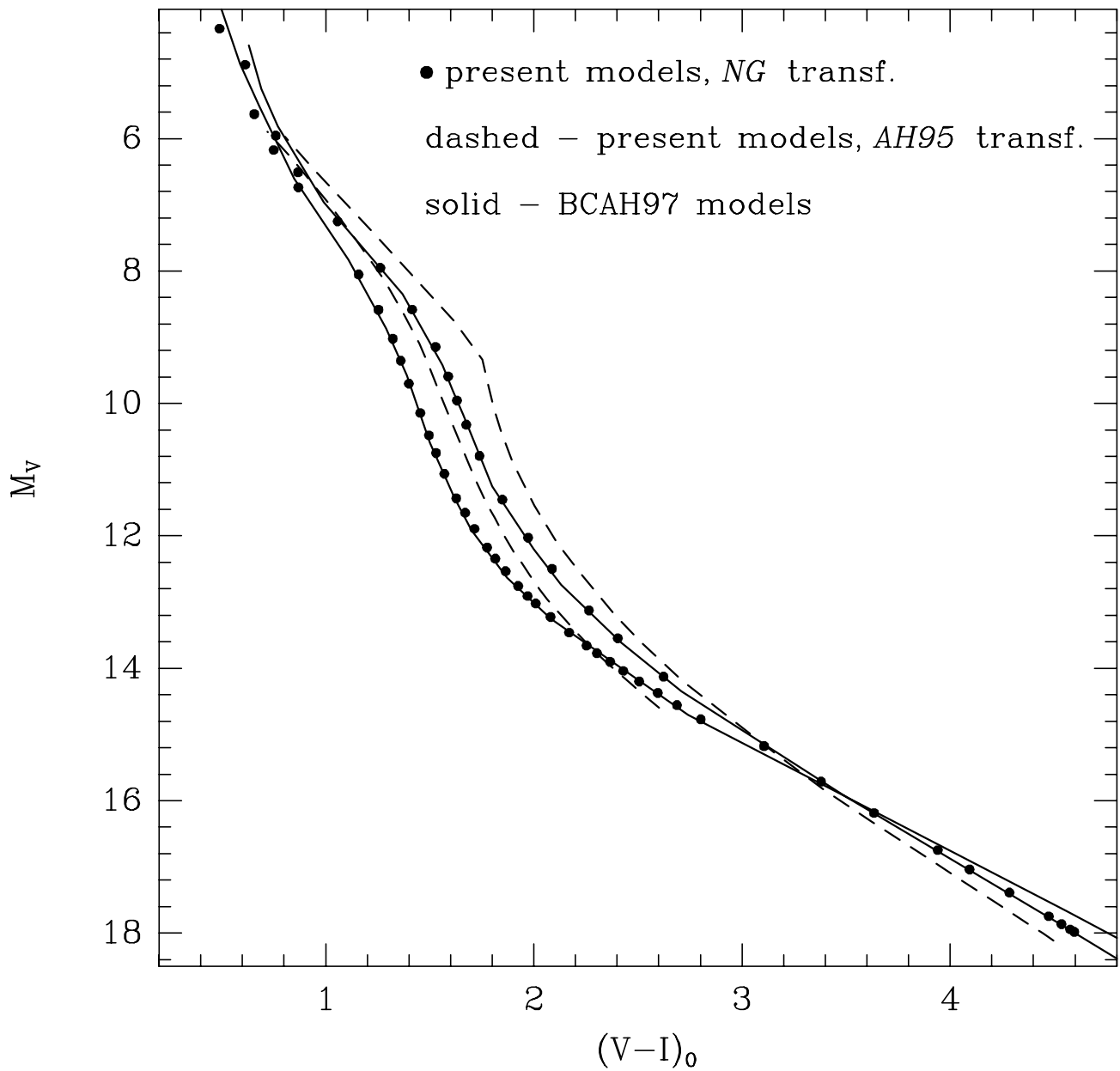
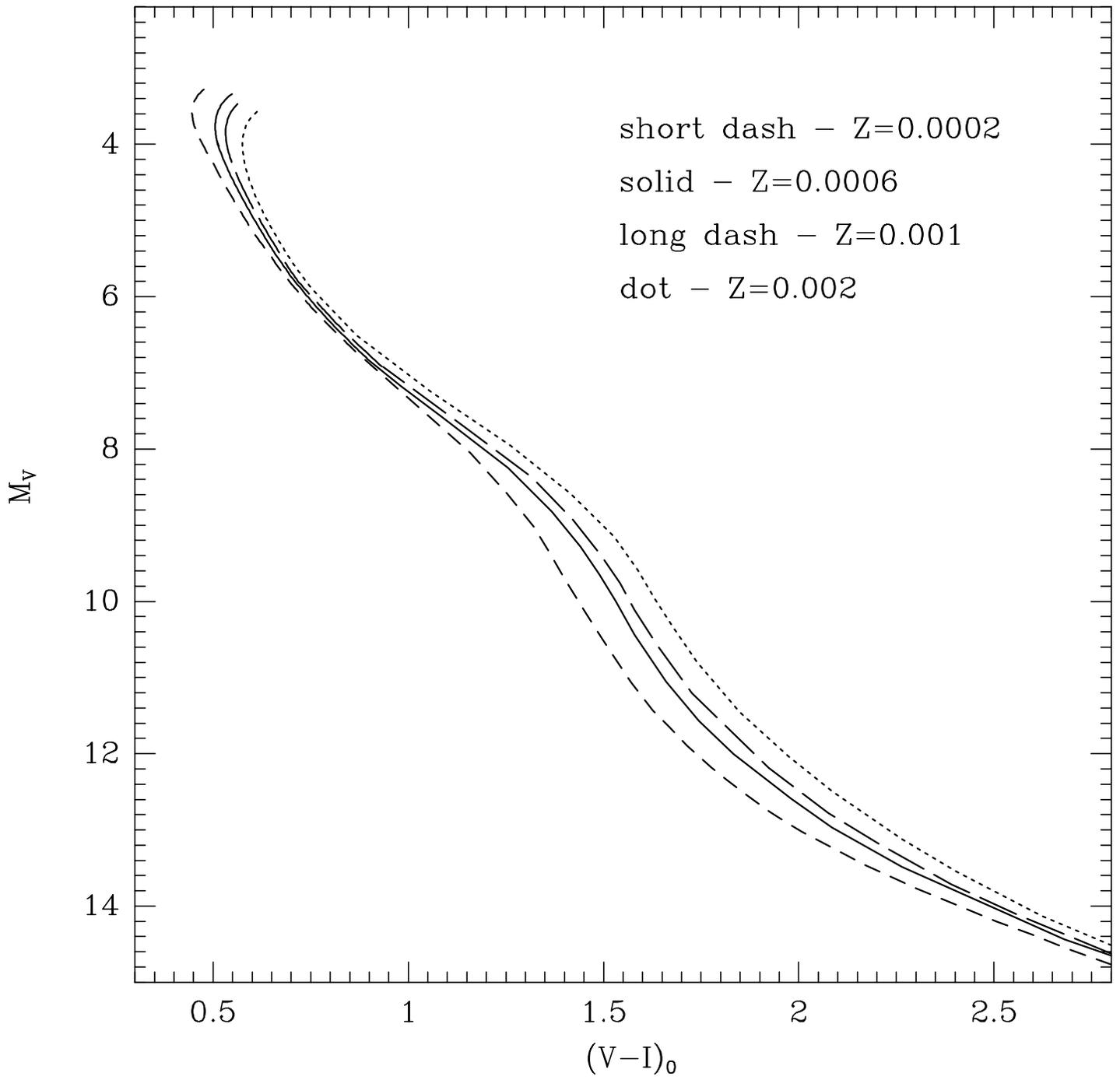
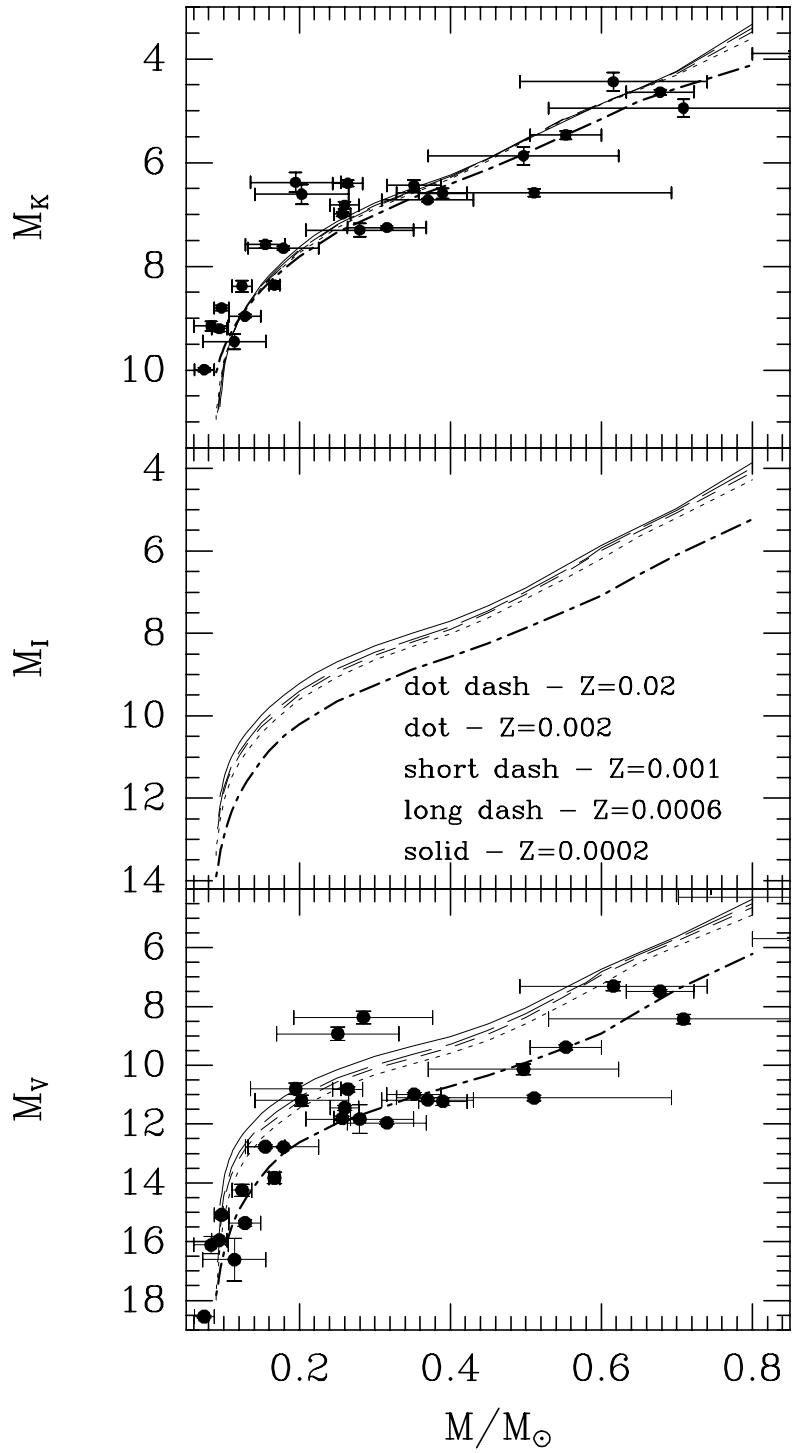
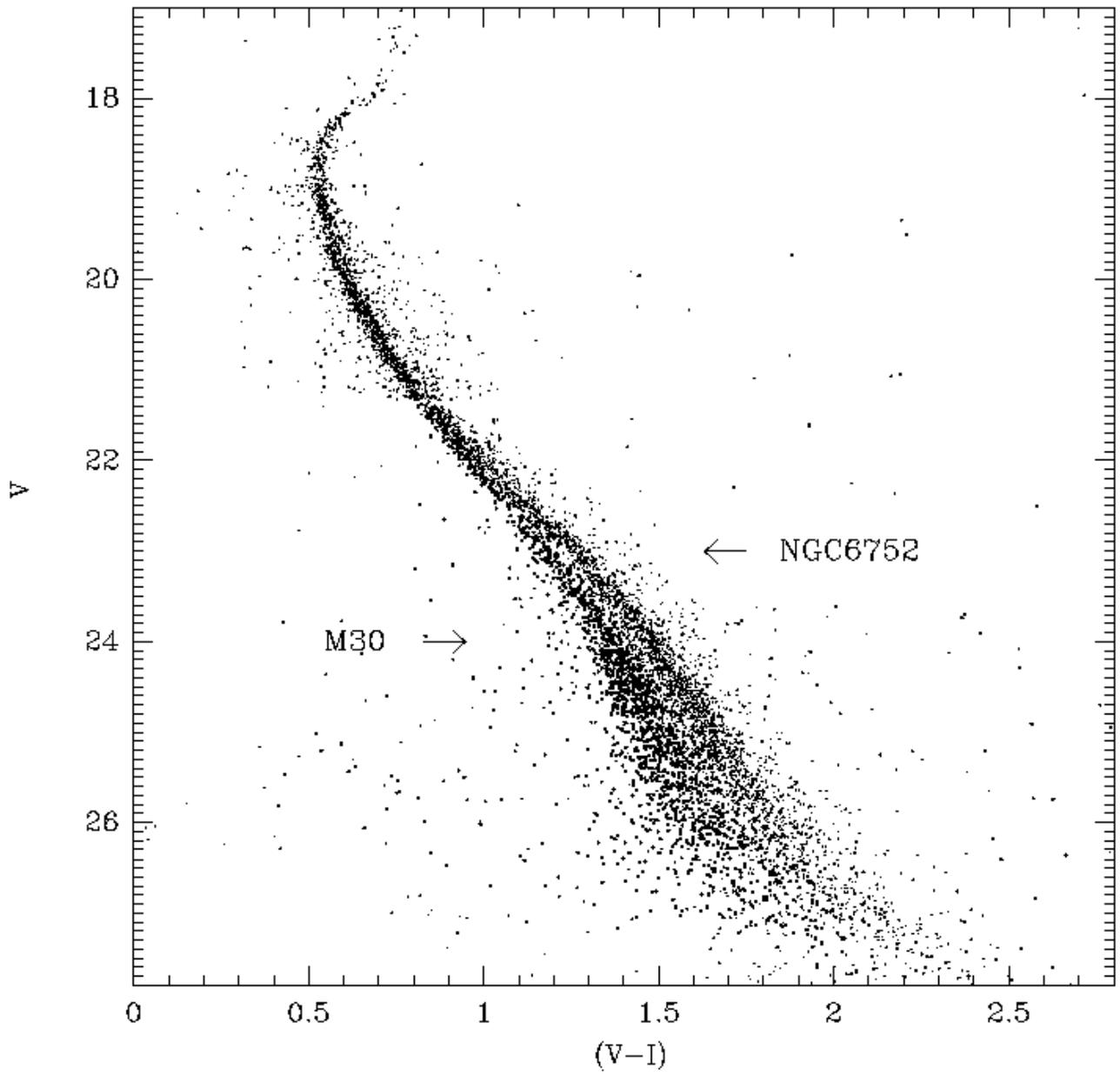
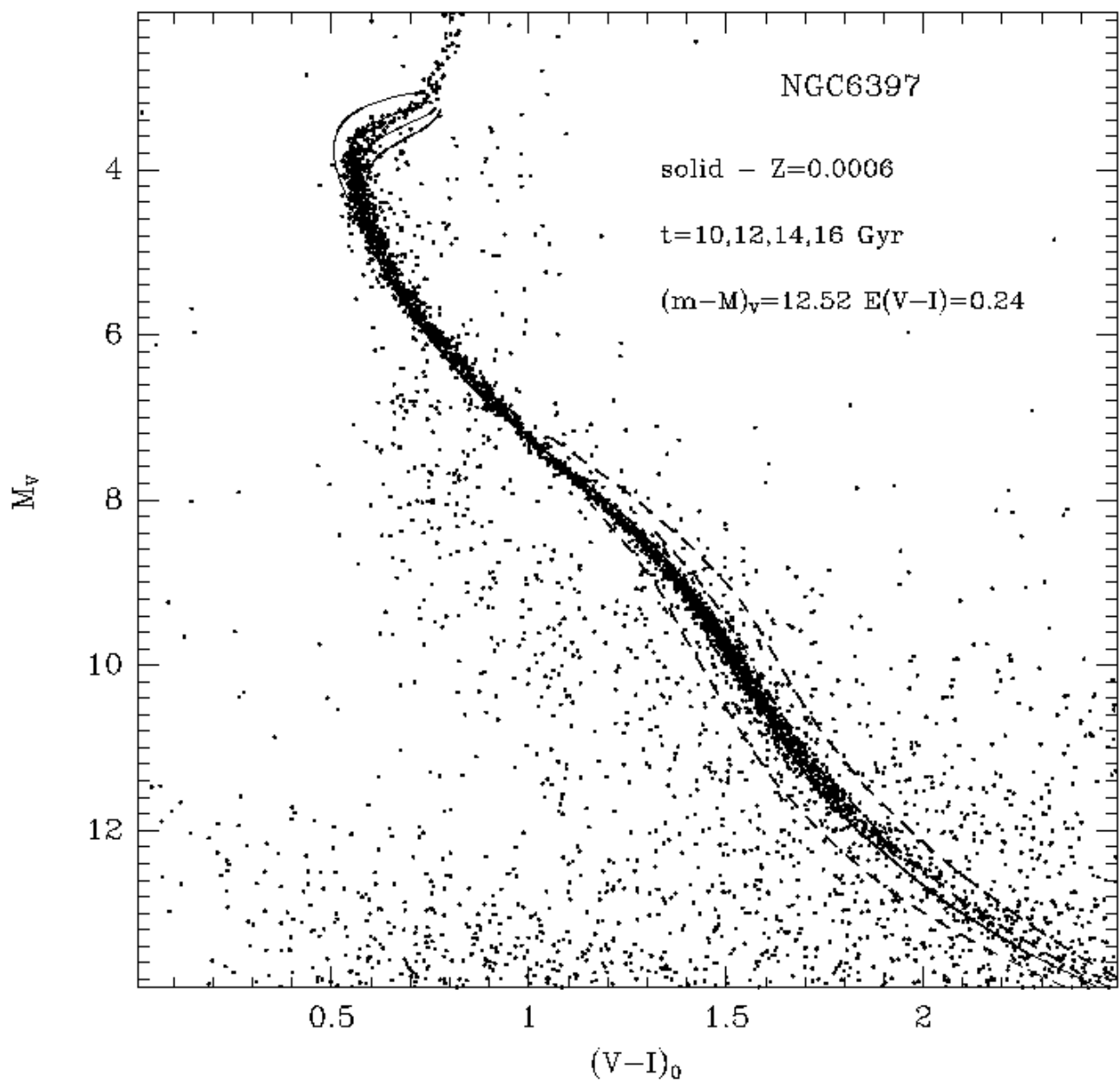
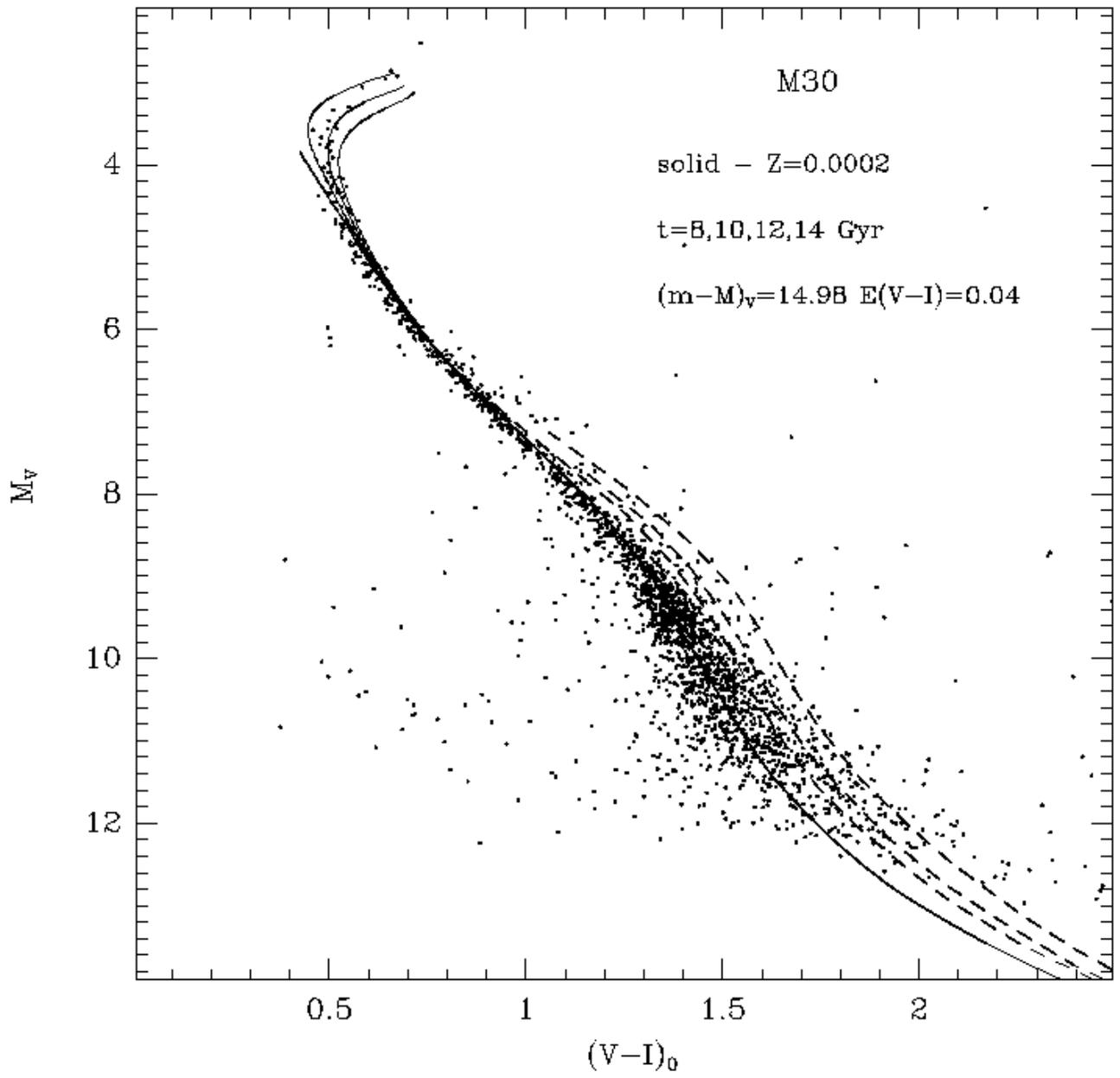
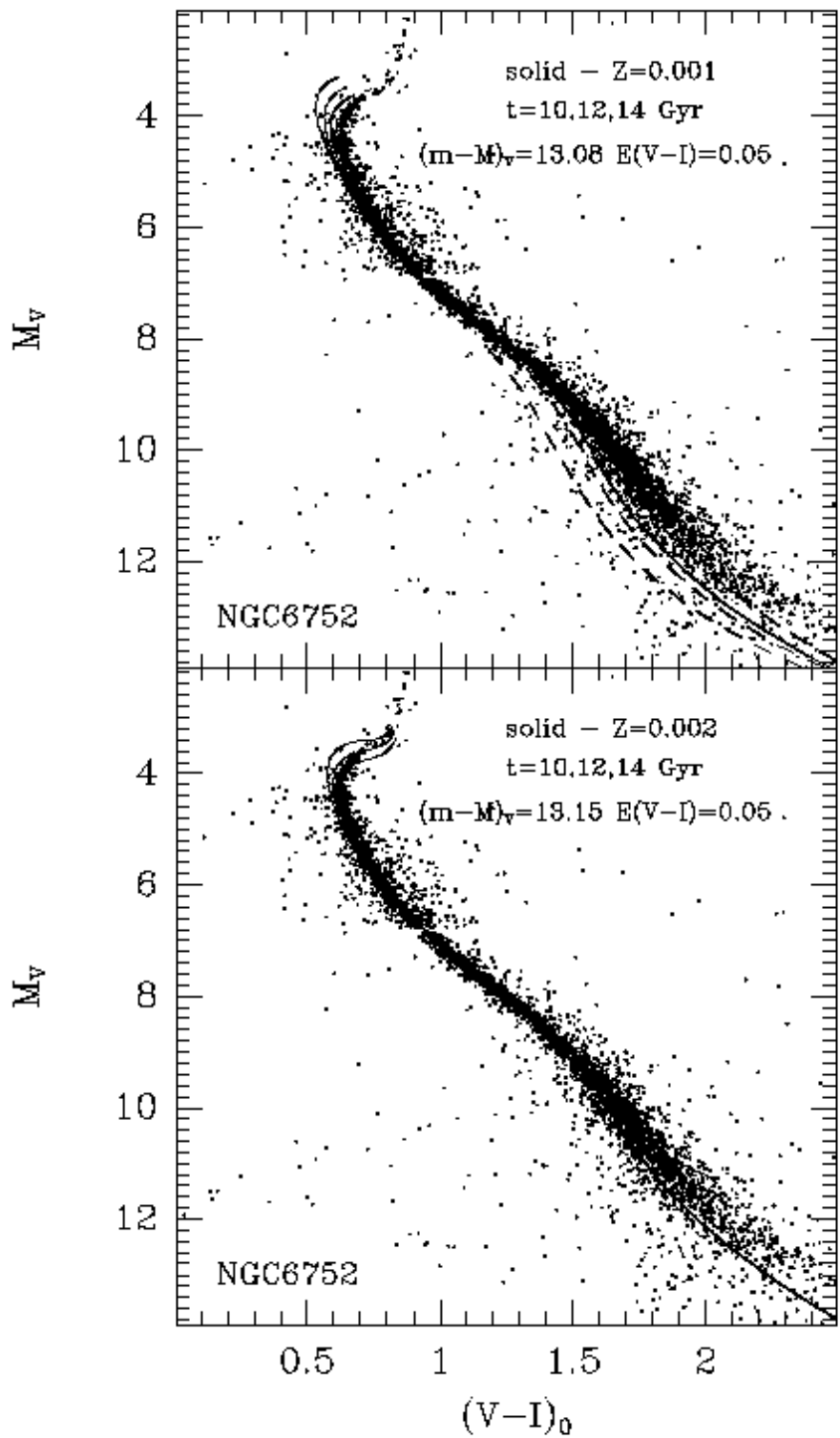
\bye